# Ponderomotive perturbations of low density low-temperature plasma under laser Thomson scattering diagnostics

Mikhail N. Shneider

*Princeton University, Princeton, NJ 08544, USA*
m.n.shneider@gmail.com

The ponderomotive perturbation in the interaction region of laser radiation with a low density and low-temperature plasma is considered. Estimates of the perturbation magnitude are determined from the plasma parameters, geometry, intensity and wavelength of laser radiation. It is shown that ponderomotive perturbations can lead to large errors in the electron density when measured using Thomson scattering.

The elastic scattering of laser radiation by free electrons (Laser Thomson Scattering - LTS) is an effective, remote non-perturbation method of plasma research [1-3]. LTS makes it possible to determine the electron density and temperature in a wide range of equilibrium and nonequilibrium plasmas - from a weakly ionized glow discharges and low-pressure discharges used in microelectronic technology [3-7], to medium- and high-pressure plasmas [8,9], as well as in plasma in tokamaks [10-12].

The total cross section of Thomson scattering by one electron [1]

$$\sigma_T = \frac{e^4}{6\pi\varepsilon_0^2 m^2 c^4} = 6.65 \cdot 10^{-29} \, \text{m}^2 \tag{1}$$

does not depend on the wavelength $\lambda$ of the laser used for diagnosis. In most practical cases of low-pressure plasmas, the parameter $\alpha = \frac{\lambda}{4\pi r_D \sin(\theta/2)} \ll 1$, therefore Thomson scattering is incoherent and the signal is proportional to the number of electrons in the focal volume $\Delta V$ from which the scattered radiation enters the photodetector [1, 2]. Here $r_D = \left(\frac{\varepsilon_0 k_B T_e}{n_e e^2}\right)^{1/2}$ is the Debye scale of screening in the plasma; $\theta$ is the angle between the observation direction and the incident laser beam. The total light flux of the scattered radiation within the limits of the solid angle $\Delta\Omega$ is determined by the expression [2]

$$\Phi_s = I_L \sigma_T n_e \Delta V \Delta\Omega, \tag{2}$$

$I_L = \frac{\varepsilon_0 E_{L,0}^2}{2} c$ is the laser beam intensity, $E_{L,0}(r)$ is the electric field amplitude in the laser beam, $\varepsilon_0$ is vacuum permittivity.

In view of the smallness of the Thomson cross section (1), LTS is effective at electron densities $n_e > 10^{11}$ cm$^{-3}$. At a reduced plasma density $n_e < 10^{11}$ cm$^{-3}$ there are very few electrons in the test volume to provide a high quality optical signal. Therefore, a large number of measurements are collected and analyzed statistically as done in [3,4,7]. In this case, observation times range from tens of minutes to several hours, making it impossible to study unsteady or turbulent plasmas. Another method is to increase the intensity of the laser radiation which may allow a reasonable number of detected photons per pulse or a substantially lower number of pulses.

However, as we show below, above an intensity threshold the LTS diagnostic ceases to be nonintrusive. We are talking about the action of the well-known ponderomotive force, which is usually not accounted for when analyzing the Thomson scattering in a low-pressure plasma at a relatively low intensity of laser radiation (<10$^{18}$ W/m$^2$). We will show that ponderomotive effects can turn out to be significant in much less intense laser radiation, especially when studying low-temperature plasma at low pressures. At the same time, the real density of electrons in the plasma can significantly differ from the one measured using LTS. To illustrate the effect, we confine ourselves to the case of a low-pressure plasma (low density) p <1 Torr, where the characteristic mean-free path of the electrons is of the order of the radius of the laser beam in the interaction region with the plasma. This will allow us to neglect the Joule heating and laser-induced avalanche ionization. On the other hand, the maximum intensity of the laser beam is selected so that multiphoton ionization, in the case of a weakly ionized plasma, cannot be neglected.

The intensity of the laser beam is nonuniform along the radius and along the axis in the region of interaction with the plasma. However, in the Rayleigh region, from which the main flux of scattered radiation is observed in LTS, the longitudinal inhomogeneity may be neglected, and only the radial inhomogeneity of a laser beam having a Gaussian radial profile can be considered:

$$I_L(r,t) = I_0(t)\exp(-r^2/r_b^2). \quad (3)$$

The ponderomotive force in the plasma acting on an electron in an inhomogeneous laser field is [13]

$$\vec{f}_p = -\frac{1}{4}\frac{e^2}{m\omega^2}\nabla E_{L,0}^2 = -\frac{1}{2}\frac{e^2}{m\omega^2 \varepsilon_0 c}\nabla I_L, \quad (4)$$

where $\omega = 2\pi c/\lambda$ is the angular frequency of the laser radiation. Electrons are displaced from the region of strong field to the periphery until the ponderomotive force is balanced by the electrostatic force in the ambipolar field

$$\vec{\nabla} \cdot \vec{E}_{amb} = \frac{1}{r}\frac{\partial(rE_{amb})}{\partial r} = \frac{e}{\varepsilon_0}(n_i - n_e) \approx \frac{e}{\varepsilon_0}\delta n, \quad (5)$$

and the resulting electron pressure gradient

$$\vec{\nabla} p_e = \vec{\nabla}(n_e k_B T_e) \approx k_B T_e \vec{\nabla} n_e \text{ if } T_e \approx \text{const} \tag{6}$$

The equilibrium condition per electron in the unperturbed plasma is

$$\vec{f}_p - e\vec{E}_{amb} - \frac{1}{n_{e,0}} \vec{\nabla} p_e = 0. \tag{7}$$

We wrote down the stationary equilibrium condition (7) because the estimate of the characteristic time of electron removal under the action of the ponderomotive force [4] in a low-pressure plasma with an electron mean-free path length of the order of the radius of the laser beam $\tau \sim (2 r_b m / |f_p|)^{1/2} < 1$ ns, which is much shorter than the typical laser pulse used in LTS [3-9], ~10 ns.

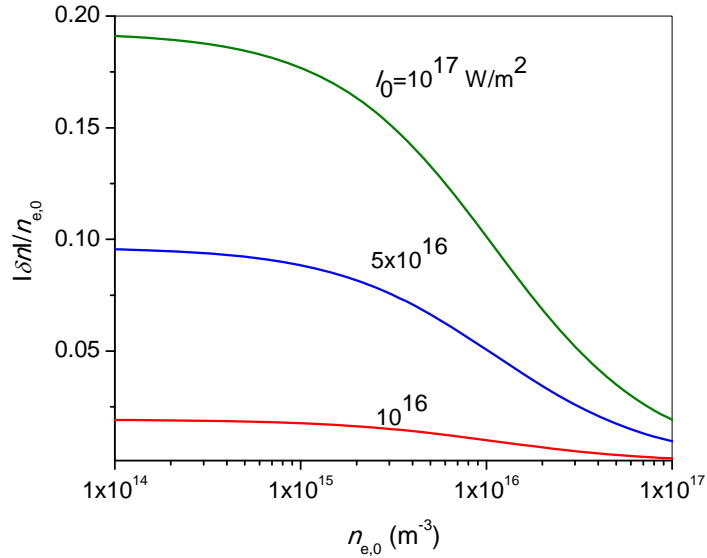

**Fig. 1.** Dependences of the relative perturbation of the plasma density in the region of LTS measurements from the unperturbed plasma density for different values of the laser beam intensity.

Taking into account that at $r = r_b$: $E_{amb} \approx \frac{r_b}{2} \frac{e}{\varepsilon_0} \delta n$, $\frac{\partial p_e}{\partial r} \approx \frac{\delta n}{r_b}$, and $\frac{\partial I_L}{\partial r} \approx -\frac{2 I_0}{r_b \overline{e}}$, $\overline{e} = 2.71828...$ is the base of natural logarithm, we obtain from (7) an estimate for the relative perturbation of the electron density in the region of interaction of the laser beam with the plasma

$$\frac{|\delta n|}{n_{e,0}} \approx \frac{e^2 I_0}{\overline{e}\omega^2 \varepsilon_0 c k_B T_e (1 + 0.5 r_b^2 / r_D^2)}. \tag{8}$$

Figure 1 shows examples of the dependences of the relative perturbation of the plasma calculated from (8) on the unperturbed electron density $n_e = n_{e,0}$ at various intensities of the laser radiation.

The results shown correspond to the parameters adopted for the example: $r_b = 100\,\mu\text{m}$, $\lambda = 532\,\text{nm}$, $T_e = 1\,\text{eV}$, which are close to those used in experiments [3-9]. When using a laser with a longer wavelength, the error increases $\propto \lambda^2$.

Thus, with increasing intensity of laser radiation used for LTS diagnostics, the error in measuring the electron density increases. This is caused by the displacement of electrons from the test area under the action of ponderomotive force. This effect must be taken into account when analyzing the results of LTS in a plasma with electron density $n_e < 10^{10}\,\text{cm}^{-3}$.

I would like to thank Drs. A. Gerakis, A. Diallo, A. Morozov and Y. Raitses for useful discussions.